# Strongly connected *ex-situ* MgB$_2$ polycrystalline bulks fabricated by solid-state self-sintering

**Hiroya Tanaka[1], Akiyasu Yamamoto[1,2], Jun-ichi Shimoyama[1], Hiraku Ogino[1] and Kohji Kishio[1]**
[1]Department of Applied Chemistry, The University of Tokyo,
7-3-1 Hongo, Bunkyo-ku, Tokyo 113-8656, Japan
[2]JST-PRESTO, 4-1-8 Honcho Kawaguchi, Saitama 332-0012, Japan

E-mail: yamamoto@appchem.t.u-tokyo.ac.jp

**Abstract**

The transport current carrying capacity of *ex-situ* processed MgB$_2$ is expected to be greatly enhanced if a strong intergrain connectivity can be realized. Although percolation theory predicts that *ex-situ* MgB$_2$ samples should have a high connectivity of over 30% due to their high bulk density ($P \approx 75\%$), the reported connectivity of *ex-situ* MgB$_2$ bulks and wires are generally less than 10%. This is presumably because *ex-situ* MgB$_2$ has a much weaker intergrain connectivity than *in-situ* MgB$_2$. It is well known that heat treatment after cold working of *ex-situ* MgB$_2$ improves the connectivity and the critical current density. However, it is currently unclear whether such heat treatment induces self-sintering that results in the formation of necks, the elimination of pores, and in increase in contact area. In the present study, we investigated the microstructure, normal-state electrical connectivity, and critical current density of *ex-situ* MgB$_2$ polycrystalline bulks prepared by systematically varying the sintering conditions under low pressure. Samples heated at a high temperature of ~900°C for a long period showed an increased packing factor, a larger intergrain contact area, and a significantly enhanced electrical connectivity, all of which indicate solid-state self-sintering of MgB$_2$. Sintered *ex-situ* MgB$_2$ bulks from a laboratory-made ball-milled powder exhibited a greatly enhanced connectivity of 28%, which is the highest connectivity of pressureless *ex-situ* MgB$_2$ bulks, wires, and tapes. Surprisingly, grain growth did not occur during long-duration (~100 h) sintering in the sintered *ex-situ* MgB$_2$ bulks. This is in marked contrast to *in-situ* processed MgB$_2$ samples for which significant grain growth occurred during heat treatment at ~900°C, producing grains that are several tens of times larger than the initial boron grains. Consequently, the critical current density as a function of the external magnetic field at 20 K progressively improved with sintering due to the relatively small grain size and good intergrain connectivity. We thus conclude that solid-state self-sintering is an effective approach for producing strongly connected, dense *ex-situ* MgB$_2$ polycrystals without grain growth.

# 1. Introduction

MgB$_2$ has the highest critical temperature $T_c$ (~40 K) [1] of intermetallic superconductors. It is expected to be applicable at liquid helium free temperatures of ~20 K, which can be easily attained by a cryocooler. MgB$_2$ bulks, wires, and tapes are reported to have critical current densities ($J_c$) at 20 K of $10^5$–$10^6$ A·cm$^{-2}$ [2–4], which is much lower than the depairing current density $J_d \approx 10^8$ A cm$^{-2}$. Indeed, epitaxial thin films have been reported to have very high $J_c$ (20 K) of ~$10^7$ A cm$^{-2}$ [5,6]. Recent studies have revealed that the transport current properties in polycrystalline MgB$_2$ samples are considerably suppressed due to poor connectivity, effective current-carrying cross-sectional area [7]. Impurity phases at grain boundaries and the low bulk density of MgB$_2$ samples are thought to be the main causes of the poor electrical connectivity of MgB$_2$ [8,9]. The *ex-situ* method that employs prereacted MgB$_2$ powder is favorable in terms of bulk density compared to the *in-situ* method. Higher bulk density of ~75% (the close-packing of rigid spheres) can be expected for the *ex-situ* method.

The *ex-situ* method is commonly used to fabricate MgB$_2$ bulks, wires, and tapes [10]. Even unsintered, as-pressed *ex-situ* processed MgB$_2$ wires and tapes exhibit a relatively high transport $J_c$ of ~$10^4$ A·cm$^{-2}$ at 20 K [11]. Heat treatment and/or impurity addition further improve $J_c$ of *ex-situ* MgB$_2$ [12–15]. However, heat-treated *ex-situ* MgB$_2$ generally does not have a higher $J_c$ than *in-situ* processed MgB$_2$, despite its higher bulk density. This is thought to be due to insufficient intergrain connectivity in *ex-situ* MgB$_2$. Indeed, moderately heat-treated *ex-situ* MgB$_2$ samples typically have a normal-state connectivity of less than 10% [16–18], which is considerably lower than those of *in-situ* MgB$_2$ [8,19–21]. This indicates that interfaces between prereacted MgB$_2$ grains/agglomerates are weakly coupled compared to the chemically formed strong coupling in *in-situ* MgB$_2$. In addition, MgB$_2$ grain surfaces are covered with oxide layers, such as MgO, BO$_x$, and amorphous oxides, which block the transport current [10,22–24].

High-temperature self-sintering is frequently used to densify polycrystalline materials such as metals, alloys, and ceramics [25]. Figure 1 shows schematic diagrams of self-sintering of three grains at four different stages. Intergrain coupling with the formation of necks between adjacent grains initially occurs (stage II). In the next stage, pores are gradually eliminated by the formation of grain boundaries. These two stages increase the surface contact area resulting in densification. However, there have been few reports of pressureless sintering of MgB$_2$ [18,26], presumably because MgB$_2$ is hard to sinter. Dancer *et al.* fabricated *ex-situ* MgB$_2$ bulks by heat treatment at 200–1100°C for 1 h and they investigated the changes in the bulk density and the amount of impurity phase [26]. They found little evidence for self-sintering as the bulk density changed by less than 3%, even for bulk samples heat treated at 1100°C. Gupta *et al.* studied the normal-state connectivity of a series of *ex-situ* MgB$_2$ bulks prepared with the addition of excess (up to 5%) Mg and for heat treatment temperatures between 700 and 950°C for 1 h [18]. They found that normal-state resistivity varied with heat-treatment temperature and the amount of excess Mg, but that particles were weakly coupled, exhibiting a connectivity of less than 7%. To date, the sintered microstructure has been achieved by using techniques that apply a high external pressure such as hot isostatic pressing [27], hot pressing [28,29], and spark plasma sintering [30,31]. On the other hand, if reasonably sintered *ex-situ* MgB$_2$ could be synthesized under low or ambient pressure, the current carrying capacity of *ex-situ* MgB$_2$ would be greatly enhanced in practical long wires.

In the present study, we investigated the microstructure, normal-state connectivity, and critical current properties of *ex-situ* MgB$_2$ polycrystalline bulks prepared by systematically varying the heating conditions under low pressure. In particular, we employed long heat treatments of up to 10 days at high temperatures of ~900°C to promote solid-state self-sintering of MgB$_2$. To prevent MgB$_2$ decomposing due to loss of Mg at high temperatures, prereacted MgB$_2$ powders were sealed and heated in a metal sheath by the powder-in-closed-tube (PICT) technique [32]. Laboratory-made MgB$_2$ powder, which has fresher surfaces than commercially available powder, was used as the starting material. We observed evidence of self-sintering of MgB$_2$ and its strong impact on the connectivity and $J_c$.

## 2. Experimental

*Ex-situ* MgB$_2$ bulks were fabricated from laboratory-made MgB$_2$ powder by the PICT method [32]. Laboratory-made MgB$_2$ powder was filled into a stainless-steel (SUS316) tube, uniaxially pressed under 500 MPa with both ends closed, and heated in an evacuated quartz ampoule. The sintering temperature and time for the *ex-situ* MgB$_2$ bulks were systematically varied from 600 to 1000°C and from 3 to 240 h, respectively. Laboratory-made MgB$_2$ powder was prepared by grinding the *in-situ* processed bulk (sintering conditions: 900°C for 2 h) synthesized from mixed powders of Mg (99.5% purity) and B (99% purity) with a molar ratio of 1:2. Grinding of the *in-situ* processed bulk MgB$_2$ was performed either by hand milling with an agate mortar or by high-energy ball milling. Ball milling was performed using a planetary ball mill with WC media for 1 h at 300 rpm. Particle size distribution analysis revealed that the ball-milled MgB$_2$ powder had a mean particle size of 2 μm, which is smaller than that (9 μm) of the hand-milled MgB$_2$ powder.

In this study, bulk samples are denoted after their starting powder, sintering temperature, and sintering time. For example, a bulk sample prepared from the hand-milled powder and sintered at 1000°C for 24 h is called "HM-1000-24" and a bulk sample synthesized from the ball-milled powder and sintered at 900°C for 24 h is "BM-900-24". The as-pressed bulk sample before sintering prepared from the hand-milled powder is termed "HM-0-0".

The constituent phases of the samples were analyzed by powder X-ray diffraction (XRD) using Cu-$K_\alpha$ radiation. Microstructural observations were performed by scanning electron microscopy (JEOL: JSM-7001F). The mass density of samples was measured by a micrometer caliper and a weighing balance. The packing factor ($P$) was determined by dividing the bulk density by the theoretical density of MgB$_2$ (2.62 g/cm$^3$). Transport measurements were performed by the conventional four-point probe method with a 15 Hz ac current using a physical property measurement system (Quantum Design: PPMS Model 6000). The normal-state electrical connectivity $K$ of samples was estimated using the equation $K = \Delta\rho_g/\Delta\rho$, where $\Delta\rho_g = \rho_g$ (300 K) $- \rho_g$ (40 K) $\equiv$ 6.32 μΩ·cm [8] and $\Delta\rho = \rho$ (300 K) $- \rho$ (40 K), which are respectively the differences in the resistivities of ideal MgB$_2$ grains and those of a sample. Magnetization was measured using a SQUID magnetometer (Quantum Design: MPMS-XL5s). $J_c$ of the bulk samples with typical dimensions of 1×1×0.5 mm$^3$ was calculated from the width of magnetization hysteresis loops based on the extended Bean model according to the equation $J_c = 20\Delta M/a(1-a/3b)$, where $\Delta M = M^+ - M^-$ is the hysteresis loop width and $a$ and $b$ ($a < b$) are the dimensions of a rectangular sample.

## 3. Results and discussion

### 3.1. Constituent phase

Figure 2 shows powder XRD patterns of the starting MgB$_2$ powders and the *ex-situ* MgB$_2$ bulks. Both the hand-milled and ball-milled MgB$_2$ starting powders are nearly single-phase MgB$_2$ with a small amount of MgO. The ball-milled powder contained a negligible amount of contamination. The ball-milled powder exhibits broader peaks than the hand-milled powder. The (110) peak of the ball-milled powder has a full-width at half-maximum (FWHM) of 0.264°, which is slightly larger than that (0.246°) of the hand-milled powder. This increase in the FWHM suggests that ball milling refines the grain size and/or introduces lattice strain [33], which agrees well with the particle size distribution analysis.

With the exception of the bulk sintered at 900°C for a long time (~240 h) that contains a trace of MgB$_4$, the sintered *ex-situ* MgB$_2$ bulks are also nearly single-phase MgB$_2$ with a small amount of MgO. MgB$_2$ is considered to decompose due to the loss of Mg through reacting with stainless-steel sheath during long sintering. In a preliminary investigation, we observed considerable evaporation of Mg and generation of MgB$_4$ when an MgB$_2$ sample was heated above ~850°C in an open system. A closed system appears to be necessary for synthesizing high-purity *ex-situ* MgB$_2$ [34], therefore the promotion of solid-state self-sintering.

Susceptibility measurements revealed that *ex-situ* MgB$_2$ bulks prepared from the hand-milled and ball-milled powders respectively had $T_c$ values of ~38.5 and ~38.1 K, which are typical for undoped MgB$_2$ samples, indicating that there was negligible contamination from the ball-milling media.

### 3.2. Bulk density and normal-state connectivity

To evaluate the influence of sintering on the bulk density, microstructure, and electrical resistivity, *ex-situ* MgB$_2$ bulks were fabricated using the hand-milled, prereacted MgB$_2$ powder by systematically varying the sintering temperature and time. Figure 3 shows the temperature dependence of the electrical resistivity for *ex-situ* bulks sintered at different temperatures (750–1000°C) but for a fixed time (24 h). The normal-state resistivity progressively decreases with increasing sintering temperature. The inset of Fig. 3 shows the sintering temperature dependence of the connectivity. The connectivity increases rapidly for samples sintered above 850°C, which suggests that the intergrain coupling is strengthened above this temperature. The connectivity was reduced for the HM-1000-24 sample. We consider that this sample has a stronger intergrain connectivity than samples sintered at lower temperatures, but that the formation of impurity phases due to the loss of Mg reduces its connectivity. Therefore, the optimum sintering temperature will be a balance between promoting intergrain coupling and the formation of impurity phases. We fixed the sintering temperature at 900°C (at which the highest connectivity was obtained) and systematically varied the sintering time.

Figure 4 shows the packing factor and the connectivity as a function of sintering time for the bulks sintered at 900°C. After heat treatment, the bulk samples shrunk relative to the as-pressed sample. The packing factor systematically improved with increasing sintering time; it reached ~75% after sintering for ~100 h (HM-900-96). The connectivity also improved with longer sintering. The HM-900-96 sample had the highest connectivity of $K = 21\%$.

*3.3. Microstructure*

Microstructural analysis was performed on samples that had been sintered to various degrees. Figure 5 shows secondary electron images of the polished surfaces of HM-0-0, HM-900-24, and HM-900-240 samples. In these images, black, gray, and white areas correspond to voids, $MgB_2$, and impurity phases, respectively. The grains in the HM-0-0 sample (Fig. 5(a)) can be individually distinguished. $MgB_2$ grains that are ~1 μm in size just contact each other at their edges, implying that the bonding between grains is relatively weak. After sintering, the HM-900-24 sample (Fig. 5(b)) had a densified structure and increased contact area between grains. Figure 5(b) shows the formation of necks (area B) and possible open pores (area C), suggesting that the HM-900-24 sample is in the middle of the self-sintering process. Local densification develops further after sintering for 240 h (Fig. 5(c)) and the agglomerate size becomes several tens of times greater than the initial grain size (~1 μm), making it difficult to distinguish the grain boundaries. Agglomeration of the impurity oxide phase (white contrast) and possible closed pores (area D) are visible inside large $MgB_2$ agglomerates, suggesting that the HM-900-240 sample is in the final stage of the self-sintering process. These images directly indicate self-sintering of $MgB_2$ during pressureless heat-treatment at 900°C.

Interestingly, the grain sizes of the sintered bulks HM-900-24 and HM-900-240 are the same as the initial grain size (~1 μm) of the HM-0-0 sample. This suggests that the grain size of *ex-situ* processed $MgB_2$ does not change much under the present sintering conditions, which is in strong contrast to *in-situ* processed $MgB_2$ where significant grain growth occurs at temperatures above 900°C [35]. We are currently performing more detailed microstructural analysis and will report the results elsewhere [36].

Figure 6(a) shows a lower-magnification secondary electron image of a polished surface of the HM-900-24 sample prepared from the hand-milled powder. Although we succeeded in strengthening the *intergrain* coupling on a micrometer scale, the *inter-agglomerate* coupling is apparently compromised on a macroscopic scale. There are dispersed islands of $MgB_2$ grains/agglomerates with sizes of 1–10 μm and voids occupy the gap between these islands. These voids apparently have different shapes from those of *in-situ* processed bulks [8] and they are much smaller. On the other hand, the $MgB_2$ grains/agglomerates have a wide size distribution. Some grains are larger than 10 μm and contain unreacted boron (dark contrast area); consequently, the sample prepared from the hand-milled powder has an inhomogeneous macroscopic structure. This inhomogeneity is considered to reduce the packing factor and the opportunity for self-sintering by reducing the contact area between grains. We thus used ball milling to improve the homogeneity of the powder size distribution of the starting powder. Figure 6(b) shows a secondary electron image of a polished surface for the BM-900-24 sample prepared from the ball-milled powder. Compared with the HM-900-24 sample (Fig. 6(a)), the particle size distribution is more uniform and the macroscopic structure is more homogeneous. A higher magnification image (not shown) shows improved intergrain coupling, as was observed in the HM-900-24 sample.

*3.4. Critical current properties and discussion*

Figure 7 shows the magnetic field dependence of $J_c$ at 20 K for *ex-situ* MgB$_2$ bulks that had been sintered to various degrees and that were prepared from different starting powders. The results for as-pressed *ex-situ* and *in-situ* processed MgB$_2$ bulks are also shown for comparison. $J_c$ systematically increased with increasing sintering time for the *ex-situ* bulks prepared from the hand-milled powder. The HM-900-48 sample has $J_c = 3.1 \times 10^5$ A/cm$^2$ at 20 K under self-field. The BM-900-48 sample prepared from the ball-milled powder showed a further enhanced $J_c$ of $4.2 \times 10^5$ A/cm$^2$, which exceeds that of our typical *in-situ* bulks (heated at 850°C for 3 h) of $\sim 3 \times 10^5$ A/cm$^2$. Figure 8 shows the sintering time dependence of the connectivity and $J_c$ (20 K, self-field) of *ex-situ* MgB$_2$ bulks prepared from the hand-milled and ball-milled powders. The connectivity improved with increasing sintering time. The bulks prepared from the ball-milled powder had a higher connectivity ($\sim 28\%$; $\Delta\rho = 22$ μΩ·cm) than those prepared from the hand-milled powder and our typical *in-situ* processed bulks heated at 850°C for 3 h ($\sim 10\%$). The dense and relatively homogeneous microstructure of the sintered bulk (Fig. 6(b)) is believed to contribute to the superior transport current path. Consequently, the bulks prepared from the ball-milled powder had higher $J_c$ values than the bulks prepared from the hand-milled powder. Here we briefly discuss the origin of excellent low-field $J_c$ observed in the sintered *ex-situ* bulks. In addition to connectivity, flux pinning strength [9] and upper critical field [37] were reported to be essential factors that determine low-field $J_c$ of *in-situ* MgB$_2$ polycrystalline samples. Since microstructural analysis showed little sign of grain growth or refinement, we may assume that the predominant flux pinning, *i.e.* grain boundary pinning, does not change much by the sintering. High field transport measurements on the present *ex-situ* bulks showed that the extrapolated $H_{c2}$ values at 20 K changes less than 5% with the progress of sintering. Therefore we conclude that the enhancement of $J_c$ in the sintered *ex-situ* MgB$_2$ bulks mainly originates from the improved packing factor and intergrain connectivity.

It is particularly noteworthy that self-sintering improved not only the self-field $J_c$ but also the irreversibility field ($H_{irr}$). The $J_c$ values of *in-situ* MgB$_2$ samples that have been sintered at high temperatures often decrease rapidly under fields. The *in-situ* bulk heated at 900°C for 24 h has a lower $H_{irr}$ than the HM-900-24 sample sintered under the same conditions. Significant grain growth during sintering (the grain size becomes several times greater than the initial boron grain size) is responsible for the deterioration of $J_c$ in *in-situ* bulks through reduction in grain boundaries (*i.e.*, flux pinning centers) [35]. In the case of the *ex-situ* MgB$_2$ bulks, long heat treatment at high temperatures ($\sim 900$°C) does not give rise to a rapid drop in $J_c$ under high fields. This implies that excessive grain growth does not occur for *ex-situ* MgB$_2$ during long heat treatment at high temperatures, which agrees well with the microstructural image shown in Fig. 5(c). This implication is further supported by flux pinning force analysis. Figure 9 shows the normalized flux pinning force $f_P = F_p/F_p^{max}$ as a function of the reduced field $h = H/H_{irr}$ at 20 K for the *ex-situ* and *in-situ* samples. Here, the flux pinning force $F_p$ is calculated by the equation $F_p = \mu_0 H J_c$, $F_p^{max}$ is the maximum value of $F_p(H)$, and $H_{irr}$ is defined as the field at which $J_c = 100$ A/cm$^2$. The $f_P(h)$ curves for the *ex-situ* MgB$_2$ bulks sintered at 900°C for 48 h prepared from the hand-milled (HM-900-48) and ball-milled (BM-900-48) powders have maximums at $h \approx 0.2$ and are similar to that of the *in-situ* bulk that did not have grain growth. In contrast, they clearly differ from that of the *in-situ* bulk heated at 900°C for 24 h with excessive grain growth. In the *in-situ* bulk heated at 900°C for 24 h that has grain sizes up to several micrometers, grain boundaries do not act as strong pinning centers and weak intragranular core pinning centers may affect its $f_P(h)$ curve at $h \approx 0.5$. These results suggest that the high-temperature sintered *ex-situ* $MgB_2$ bulks have a similar flux pinning mechanism as typical *in-situ* $MgB_2$ bulks that do not have grain growth.

The detailed self-sintering mechanism and the reason why significant grain growth does not occur in the present *ex-situ* $MgB_2$ bulks are currently unclear. Thermodynamic calculations predict that $MgB_2$ will melt at 2430°C (~2700 K) at pressures above 49,000 Torr [38], while under ambient pressure (760 Torr) $MgB_2$ decomposes above 1545°C into a mixture of $MgB_4$ and Mg vapor. The sintering temperature ($T_{sinter}$ = 900°C) employed in the present study is significantly lower than the melting point of $MgB_2$. The ratio $T_{sinter}/T_{melt} \approx 0.43$ is considerably lower than that for sintering of ceramics for which $T_{sinter}/T_{melt}$ is typically greater than ~0.7 [25]. The sintering temperature of $T_{sinter} \approx 0.43 T_{melt}$ is considered to be long on promoting self-sintering but short on rearranging grain boundaries, thus grain growth. In addition, impurity phases at grain boundaries may hinder grain growth. Detailed microstructural analysis should be performed to further increase the packing density of sintered *ex-situ* $MgB_2$ and optimize their critical current properties.

## 4. Conclusions

We have synthesized *ex-situ* $MgB_2$ polycrystalline bulks prepared by systematically varying the heating conditions under ambient pressure to study the effects of self-sintering on the microstructure, connectivity, and $J_c$ of $MgB_2$. We found that high-temperature (~900°C) heat treatment in a closed system promoted self-sintering of *ex-situ* $MgB_2$ bulks. With prolonged heat-treatment, further increases were observed in the intergrain contact area, the bulk density, and the electrical connectivity. The sintered *ex-situ* $MgB_2$ bulks showed a high connectivity of $K$ = 21%, which is comparable to or higher than those of typical *in-situ* $MgB_2$ bulks. Control of the grain size distribution of the initial $MgB_2$ powders by ball milling was found to improve the microstructural homogeneity and promote self-sintering. Sintered *ex-situ* $MgB_2$ bulks prepared from the ball-milled powder had an enhanced connectivity of $K$ = 28%. This excellent connectivity is the highest reported connectivity for pressureless *ex-situ* $MgB_2$ and it is approaching that of ($K \approx$ 30–40%) hot-isostatic-pressed $MgB_2$. Interestingly, $MgB_2$ grains in the sintered *ex-situ* bulks were the same size as the initial grains. Strong intergrain coupling together with the small grain size contributed to the high $J_c$ and $J_c$ = $4.2 \times 10^5$ A/cm$^2$ at 20 K was recorded. The present results suggest that moderately high-temperature long heat-treatment is a simple but highly effective method for extracting the current carrying potential of high bulk density *ex-situ* $MgB_2$ through enhancing intergrain coupling by self-sintering.

**Acknowledgement**

The authors gratefully acknowledge S. Ohtsuka and T. Moroyama (Univ. of Tokyo) for experimental assistance with the scanning electron microscopy analysis and Y. Shimada and S. Hata (Kyushu Univ.), and T. Matsushita (Kyushu Institute of Technology) for fruitful discussions. This work was partially supported by the Japan Science and Technology Agency, PRESTO, and by Grant-in-Aid for Scientific Research from the Japan Society for the Promotion of Science (Grant Nos. 22860019, 23246110, and 24656368). Microstructural analysis was conducted at the Center for Nano Lithography & Analysis, The University of Tokyo, supported by the Ministry of Education, Sports, Culture, Science and Technology (MEXT), Japan.

**References**

[1] Nagamatsu J, Nakagawa N, Muranaka T, Zenitani Y and Akimitsu J 2001 *Nature* **410** 63

[2] Larbalestier D C *et al.* 2001 *Nature* **410** 186

[3] Collings E W, Sumption M D, Bhatia M, Susner M A and Bohnenstiehl S D 2008 *Supercond. Sci. Technol.* **21** 103001

[4] Eisterer M 2007 *Supercond. Sci. Technol.* **20** R47

[5] Zhuang C G, Meng S, Zhang C Y, Feng Q R, Gan Z Z, Yang H, Jia Y, Wen H H and Xi X X 2008 *J. Appl. Phys.* **104** 013924

[6] Naito M, Yamamoto A, Ueda S and Nishiyuki K 2011 *Appl. Phys. Express* **4** 073101

[7] Rowell J M 2003 *Supercond. Sci. Technol.* **16** R17

[8] Yamamoto A, Shimoyama J, Kishio K and Matsushita T 2007 *Supercond. Sci. Technol.* **20** 658

[9] Matsushita T, Kiuchi M, Yamamoto A, Shimoyama J and Kishio K 2008 *Supercond. Sci. Technol.* **21** 015008

[10] Braccini V, Nardelli D, Penco R and Grasso G 2007 *Physica C* **456** 209

[11] Grasso G, Malagoli A, Ferdeghini C, Roncallo S, Braccini V, Siri A S and Cimberle M R 2001 *Appl. Phys. Lett.* **79** 230

[12] Dhalle M, Toulemonde P, Beneduce C, Musolino N, Decroux M and Flukiger R 2001 *Physica C* **363** 155

[13] Tachikawa K, Yamada Y, Suzuki O, Enomoto M and Aodai M 2002 *Physica C* **382** 108

[14] Malagoli A, Grasso G, Tumino A, Modica M, Braccini V, Roncallo S, Bellingeri E, Ferdeghini C and Siri A S 2003 *Inter. J. Mod. Phys.* **B 23** 461

[15] Kovac P, Melisek T, Kopera L, Husek I, Polak M and Kulich M 2009 *Supercond. Sci. Technol.* **22** 075026

[16] Malagoli A, Braccini V, Tropeano M, Vignolo M, Bernini C, Fanciulli C, Romano G, Putti M, Ferdeghini C, Mossang E, Polyankii A and Larbalestier D C 2008 *J. Appl. Phys.* **104** 103908

[17] Nakane T and Kumakura H 2009 *IEEE Trans. Appl. Supercond.* **19** 2793

[18] Gupta A, Kumar A and Narlikar A V 2009 *Supercond. Sci. Technol.* **22** 105005

[19] Matsumoto A, Kumakura H, Kitaguchi H, Senkowicz B J, Jewell M C, Hellstrom E E, Zhu Y, Voyles P M and Larbalestier D C 2006 *Appl. Phys. Lett.* **89** 132508

[20] Kim J H, Oh S, Heo Y U, Hata S, Kumakura H, Matsumoto A, Mitsuhara M, Choi S, Shimada Y, Maeda M, MacManus-Driscoll J L and Dou S X 2012 NPG Asia Mater. **4** e3; doi:10.1038/am.2012.3

[21] Wang C D, Ma Y W, Zhang X P, Wang D L, Gao Z S, Yao C, Wang C L, Oguro H, Awaji S and Watanabe K 2012 *Supercond. Sci. Technol.* **25** 035018

[22] Fujii H, Kumakura H and Togano K 2001 *Physica C* **363** 237

[23] Jiang J, Senkowicz B J, Larbalestier D C and Hellstrom E E 2006 *Supercond. Sci. Technol.* **19** L33

[24] Kovac P, Husek I, Melisek T, Grivel J C, Pachla W, Strbik V, Diduszko R, Homeyer J and Andersen N H 2004 *Supercond. Sci. Technol.* **17** L41

[25] for example, German R M 1996 *Sintering Theory and Practice* (New York, NY: John Wiley & Sons,

Inc.)

[26] Dancer C E J, Mikheenko P, Bevan A, Abell J S, Todd R I and Grovenor C R M 2009 *J. Euro. Ceram. Soc.* **29** 1817

[27] Senkowicz B J, Mungall R J, Zhu Y, Jiang J, Voyles P M, Hellstrom E E and Larbalestier D C 2008 *Supercond. Sci. Techonol.* **21** 35009

[28] Tampieri A, Celotti G, Sprio S, Caciuffo R and Rinaldi D 2004 *Physica C* **400** 97

[29] Kovac P, Husek I, Melisek T, Fedor J, Cambel V, Morawski A and Kario A 2009 *Physica C* **469** 713

[30] Song K J, Park C, Kim S W, Ko R K, Ha H S, Kim H S, Oh S S, Kwon Y K, Moon S H and Yoo S I 2005 *Physica C* **426** 588

[31] Dancer C E J, Prabhakaran D, Basoglu M, Yanmaz E, Yan H, Reece M, Todd R I and Grovenor C R M 2009 *Supercond. Sci. Techonol.* **22** 095003

[32] Yamamoto A, Shimoyama J, Ueda S, Katsura Y, Horii S and Kishio K 2004 *Supercond. Sci. Technol.* **17** 921

[33] Yamamoto A, Shimoyama J, Ueda S, Katsura Y, Iwayama I, Horii S and Kishio K 2005 *Appl. Phys. Lett*. **86** 212502

[34] Yamamoto A, Tanaka H, Shimoyama J, Ogino H, Kishio K and Matsushita T 2012 *Jpn. J. Appl. Phys.* **51** 010105

[35] Yamamoto A, Shimoyama J, Ueda S, Katsura Y, Iwayama I, Horii S and Kishio K 2005 *Physica C* **426** 1220

[36] Shimada Y *et al.*, in preparation

[37] Kim J H, Oh S, Kumakura H, Matsumoto A, Heo Y U, Song K S, Kang Y M, Maeda M, Rindfleisch M, Tomsic M, Choi S and Dou S X 2011 *Adv. Mater*. **23** 4942

[38] Liu Z K, Schlom D G, Li Q and Xi X X 2001 *Appl. Phys. Lett*. **78** 3678

**Figure captions:**

**Figure 1.** Schematic diagrams of stages in solid-state self-sintering. Stage I: before sintering. Stage II: early stages of sintering; contact angle of grains becomes shallower (formation of necks). Stage III: middle stage of sintering; further evolution of necks and pores diffuse through and accumulate at grain boundaries (GBs). Stage IV: final stage of self-sintering; some pores are eliminated and isolated and closed pores appear. The contact GB area between grains progressively increases and densification occurs during sintering.

**Figure 2.** XRD patterns of starting $MgB_2$ powders ground by (a) hand milling (HM powder) and (b) ball milling (BM powder) and *ex-situ* $MgB_2$ bulks (c) HM-900-24, (d) HM-900-240, and (e) BM-900-24.

**Figure 3.** Temperature dependence of electrical resistivity $\rho$ of *ex-situ* bulks prepared from hand-milled powder (HM) and sintered for 24 h at different temperatures. The inset shows sintering temperature $T_{sinter}$ dependence of connectivity $K$.

**Figure 4.** Packing factor $P$ and connectivity $K$ as a function of sintering time $t$ at 900°C for *ex-situ* bulks obtained from hand-milled powder (HM).

**Figure 5.** Secondary electron images of cross-sectional polished surfaces of *ex-situ* $MgB_2$ bulks; (a) as-pressed bulk before sintering (HM-0-0), (b) bulk sintered at 900°C for 24 h (HM-900-24), and (c) bulk sintered at 900°C for 240 h (HM-900-240). The inset of (c) shows a magnified image of the region indicated by the white dotted square in (c). Areas A, B, C, and D indicate characteristic microstructural regions produced during self-sintering. Area A in (a) indicate "egg" of necks. Areas B and C in (b) indicate a neck and an open pore, respectively. Area D in (c) indicate closed pores.

**Figure 6.** Secondary electron images of cross-sectional polished surfaces of *ex-situ* $MgB_2$ bulks sintered at 900°C for 24 h prepared from (a) hand-milled powder (HM-900-24) and (b) ball-milled powder (BM-900-24).

**Figure 7.** Magnetic field dependence of $J_c$ at 20 K for *ex-situ* $MgB_2$ bulks (HM-0-0, HM-900-24, HM-900-48, and BM-900-48). Data for *in-situ* bulks heat treated at 900°C for 24 h and at 850°C for 3 h are shown for comparison.

**Figure 8.** Sintering time $t$ dependence of connectivity $K$ and $J_c$ at 20 K under self-field for *ex-situ* $MgB_2$ bulks sintered at 900°C prepared from hand-milled (HM) and ball-milled (BM) powders.

**Figure 9.** Normalized flux pinning force $f_p = F_p/F_p^{max}$ as a function of reduced field $h = H/H_{irr}$ at 20 K for the *ex-situ* $MgB_2$ bulks sintered at 900°C for 48 h prepared from hand-milled powder (HM-900-48) and ball milled powder (BM-900-48). Data for *in-situ* bulks heat-treated at 850°C for 3 h and at 900°C for 24 h are shown for comparison.

**Figure 1**

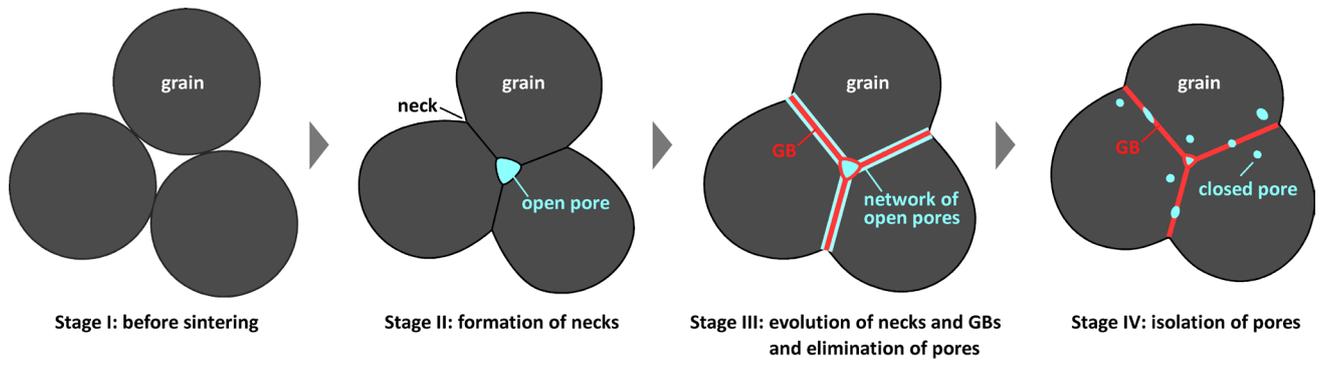

Stage I: before sintering   Stage II: formation of necks   Stage III: evolution of necks and GBs and elimination of pores   Stage IV: isolation of pores

Figure 2

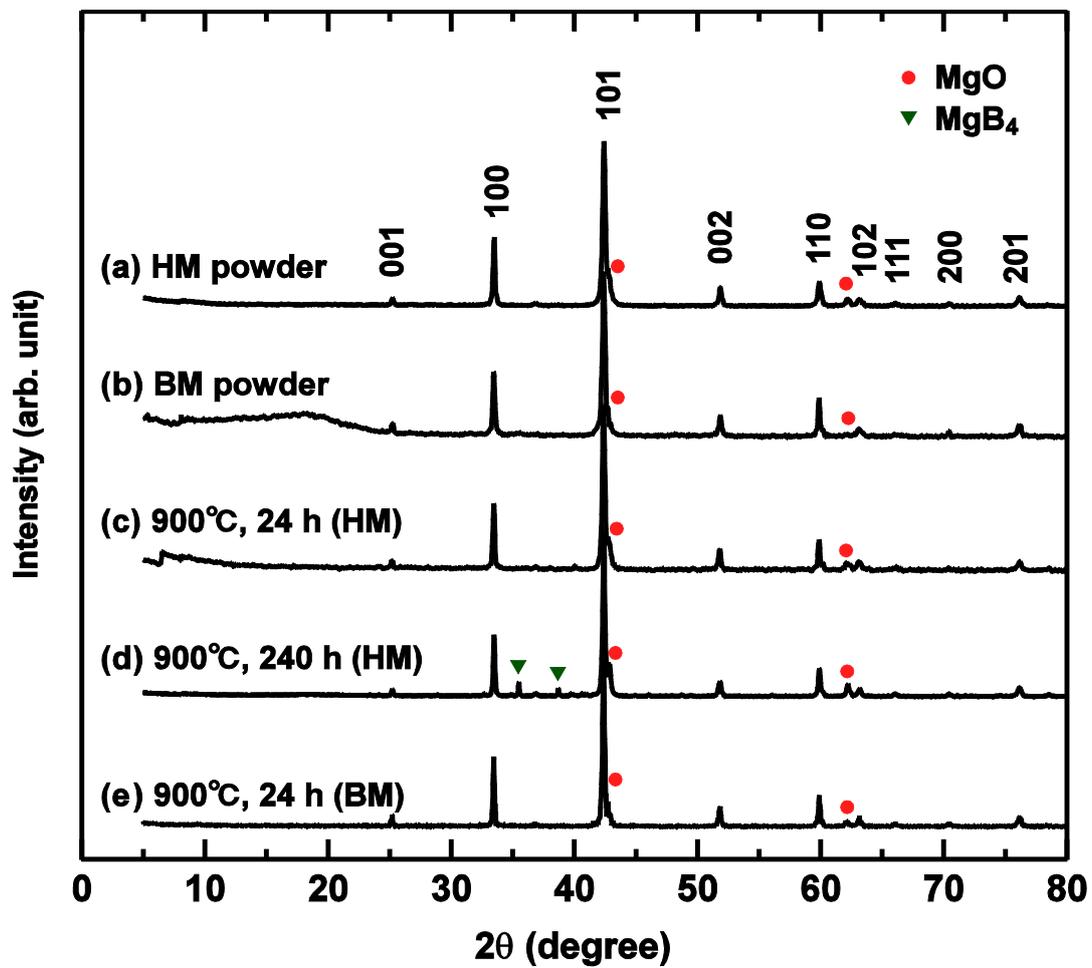

Figure 3

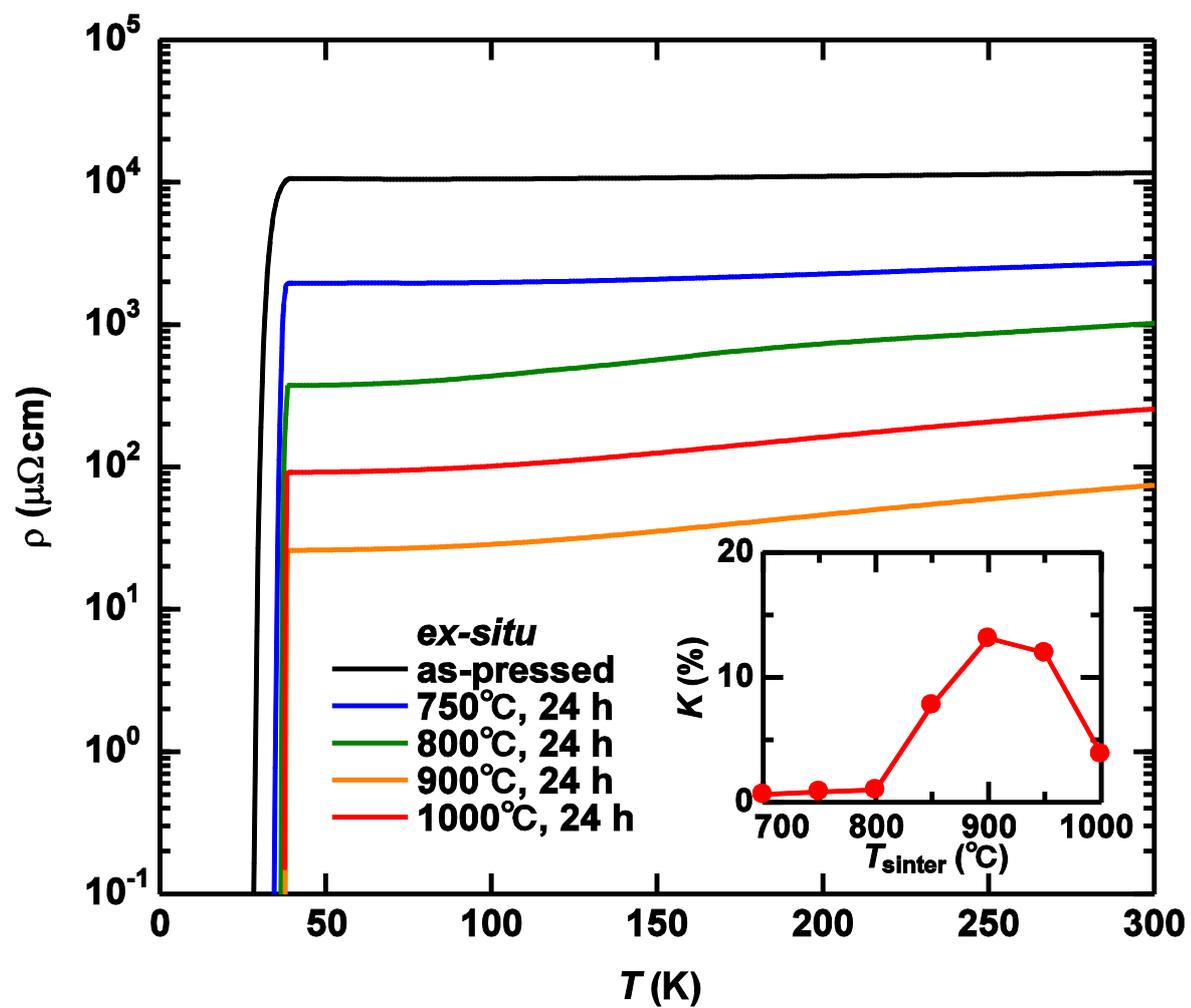

**Figure 4**

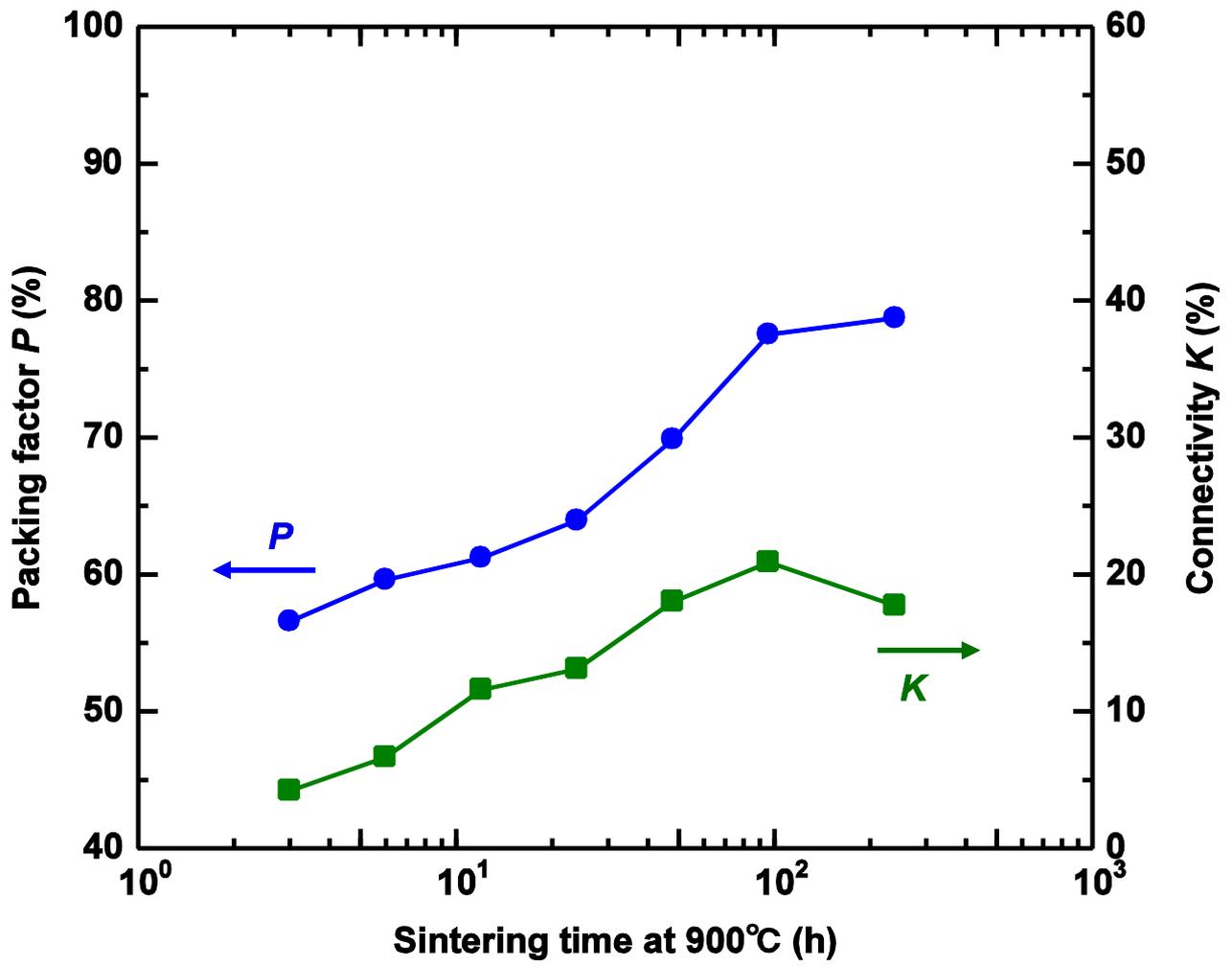

**Figure 5**

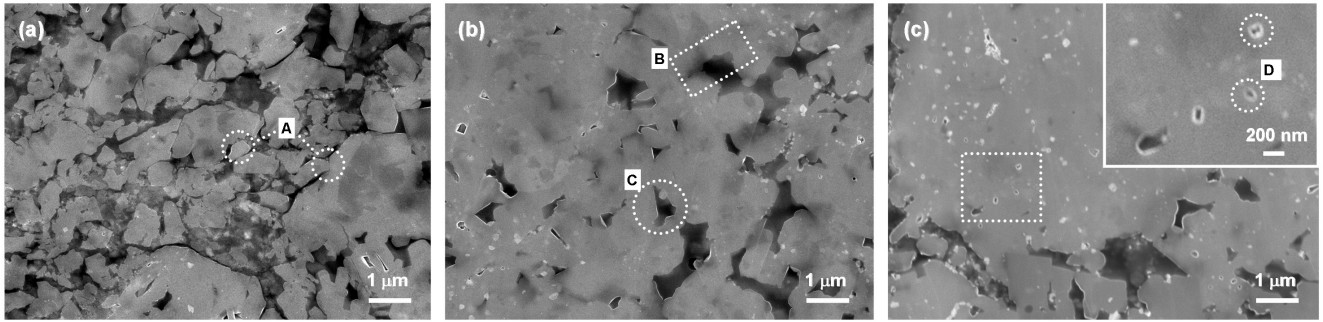

**Figure 6**

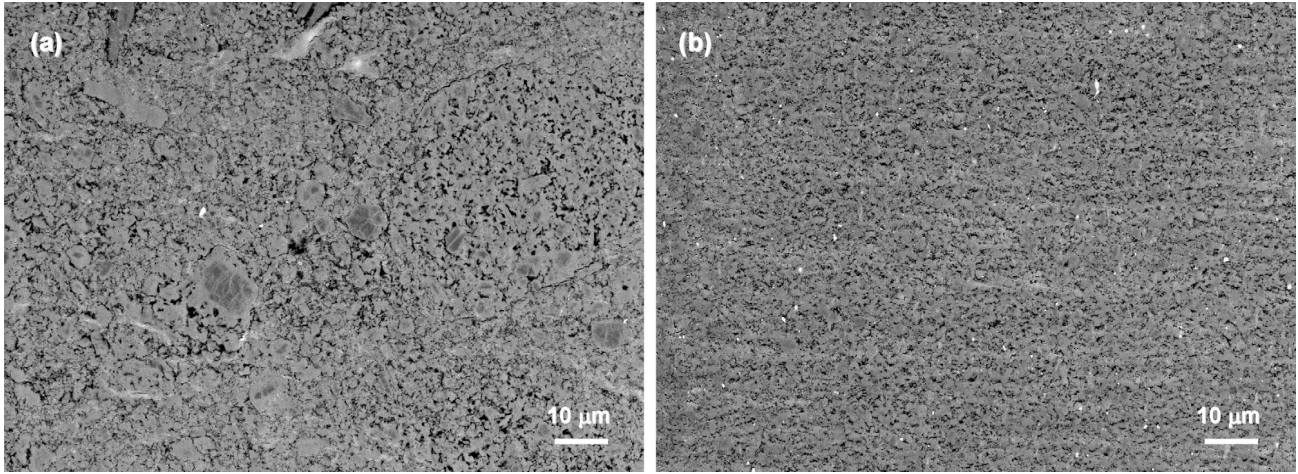

Figure 7

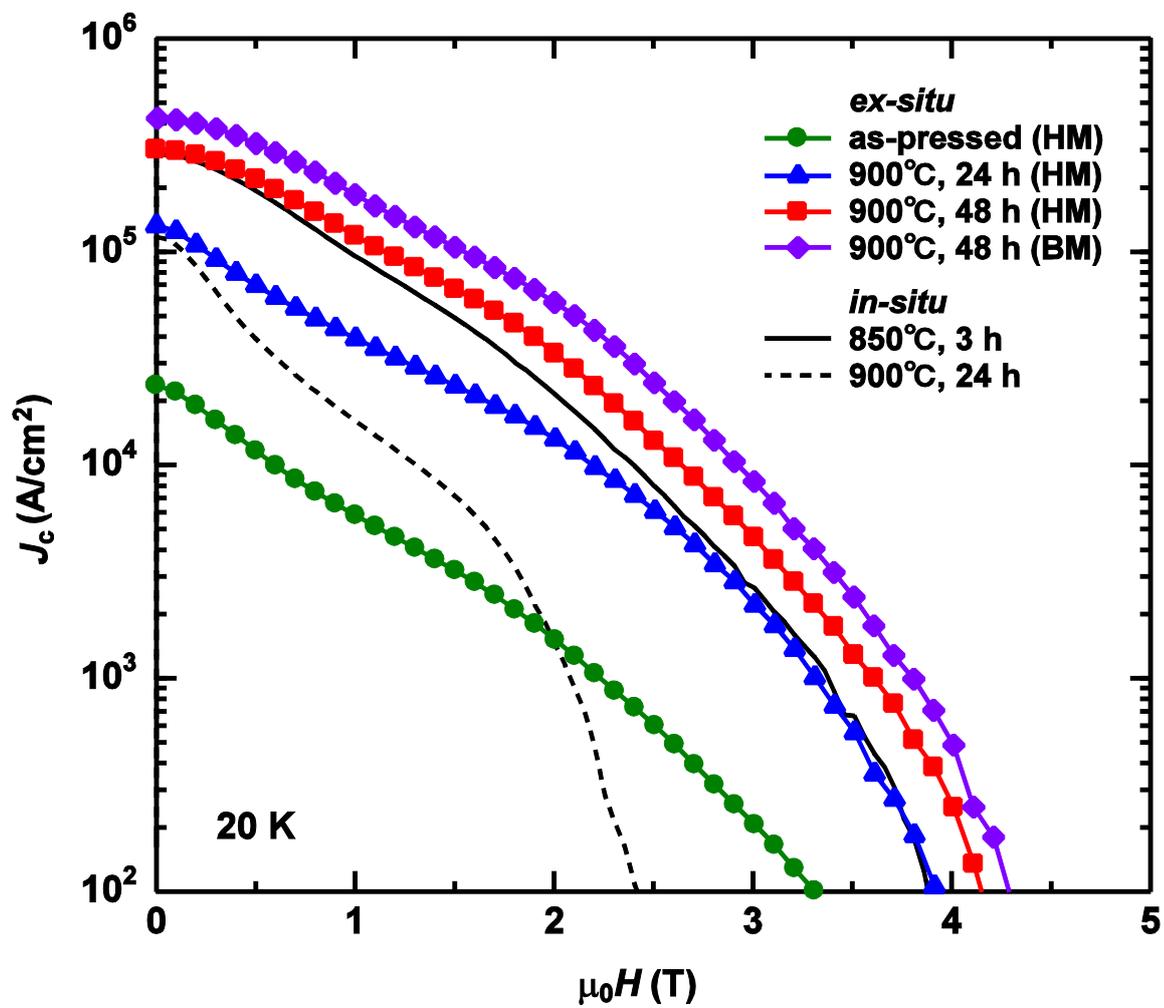

Figure 8

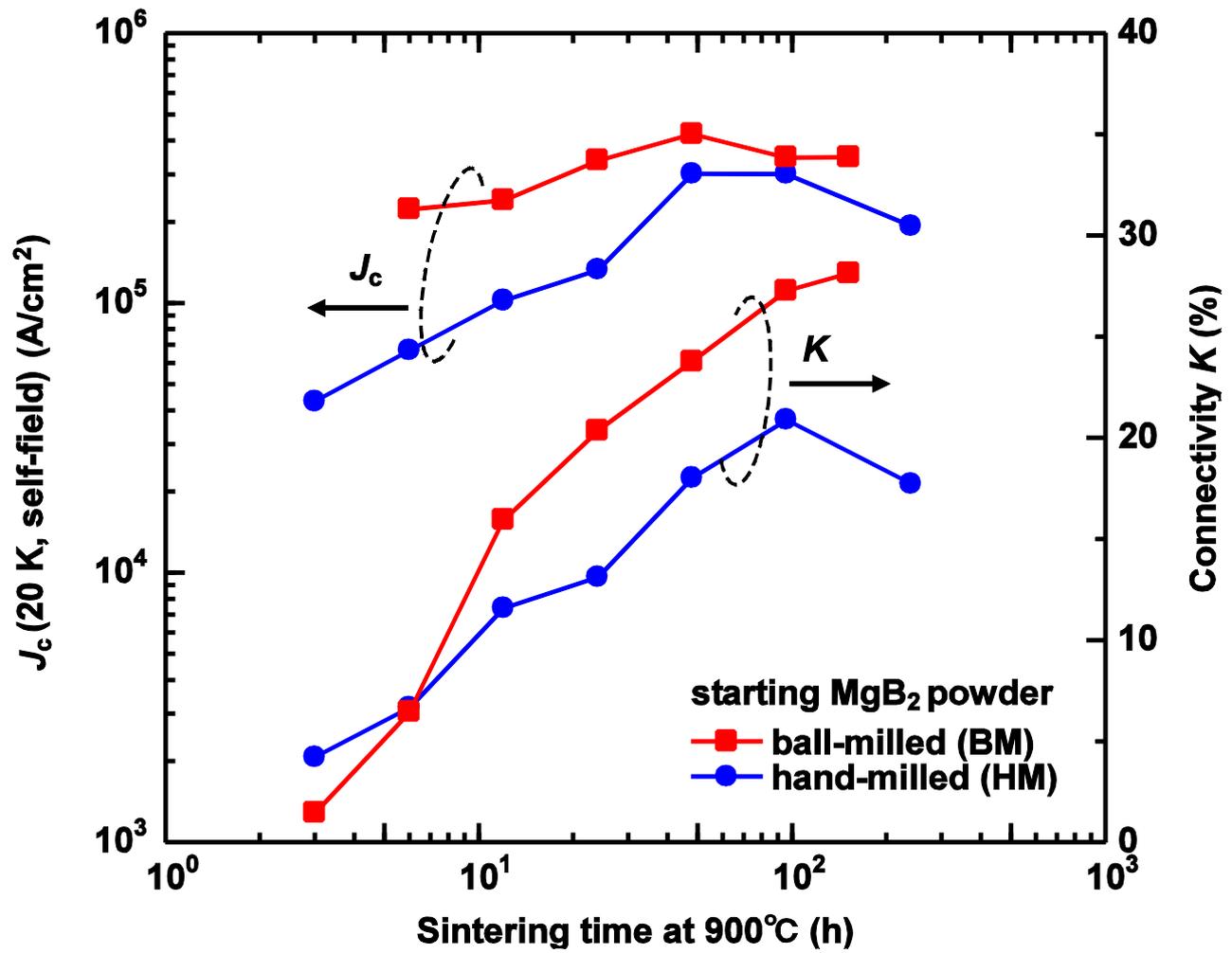

Figure 9

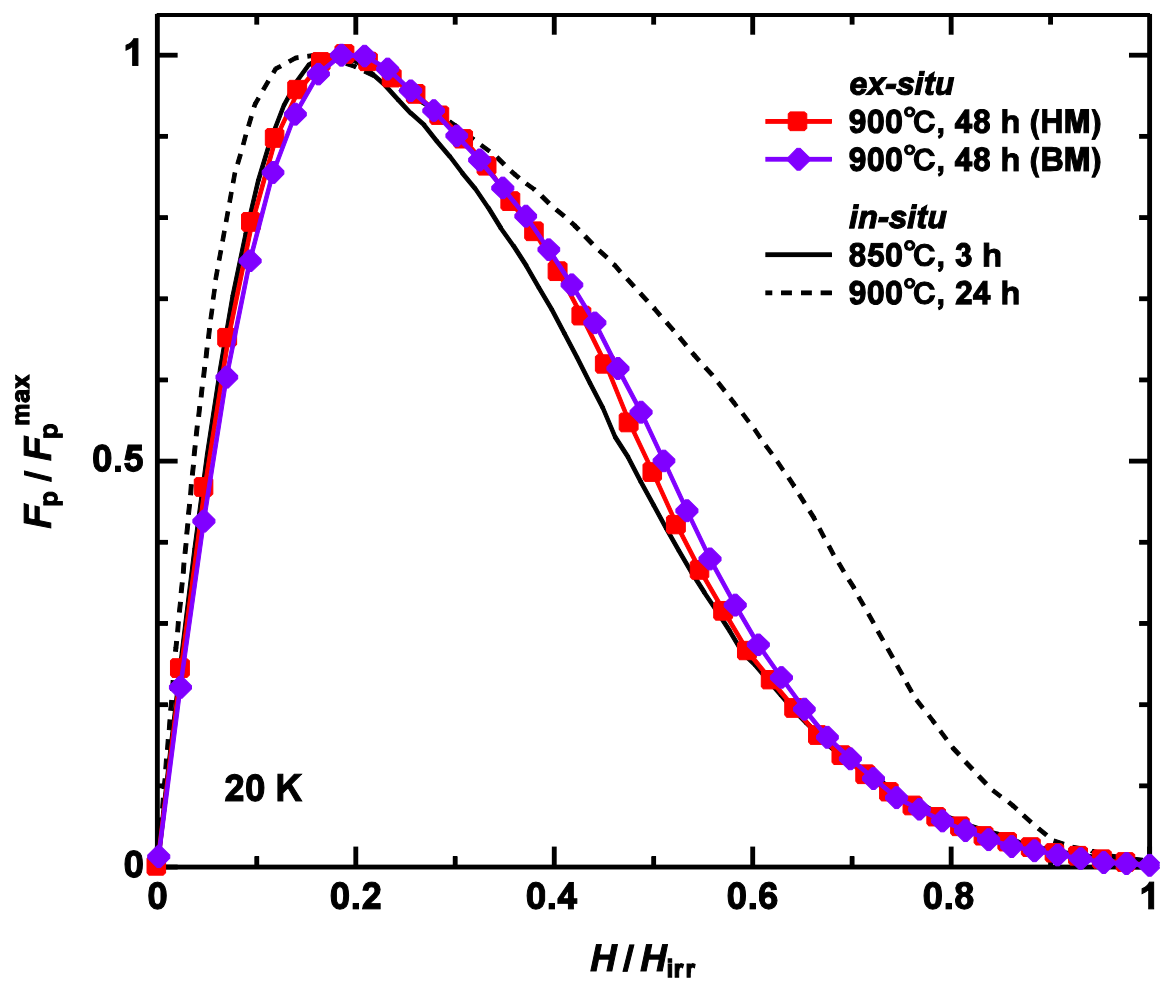

21